\documentclass[sigconf]{acmart}
\acmConference[ESEC/FSE 2022]{The 30th ACM Joint European Software Engineering Conference and Symposium on the Foundations of Software Engineering}{14 - 18 November, 2022}{Singapore}
\AtBeginDocument{%
  \providecommand\BibTeX{{%
    \normalfont B\kern-0.5em{\scshape i\kern-0.25em b}\kern-0.8em\TeX}}}

\setcopyright{acmcopyright}
\copyrightyear{2018}
\acmYear{2018}
\acmDOI{XXXXXXX.XXXXXXX}

\acmConference[Conference acronym 'XX]{Make sure to enter the correct
  conference title from your rights confirmation emai}{June 03--05,
  2018}{Woodstock, NY}
\acmBooktitle{Woodstock '18: ACM Symposium on Neural Gaze Detection,
 June 03--05, 2018, Woodstock, NY} 
\acmPrice{15.00}
\acmISBN{978-1-4503-XXXX-X/18/06}

\usepackage{listings}
\usepackage{xcolor}
\definecolor{codegreen}{rgb}{0,0.6,0}
\definecolor{codegray}{rgb}{0.5,0.5,0.5}
\definecolor{codepurple}{rgb}{0.58,0,0.82}
\definecolor{backcolour}{rgb}{0.95,0.95,0.92}

\lstdefinestyle{mystyle}{
    backgroundcolor=\color{backcolour},   
    commentstyle=\color{codegreen},
    keywordstyle=\color{magenta},
    numberstyle=\tiny\color{codegray},
    stringstyle=\color{codepurple},
    basicstyle=\ttfamily\footnotesize,
    breakatwhitespace=false,         
    breaklines=true,                 
    captionpos=b,                    
    keepspaces=true,                 
    numbers=left,                    
    numbersep=5pt,                  
    showspaces=false,                
    showstringspaces=false,
    showtabs=false,                  
    tabsize=2
}
\begin{document}

\title{TAPHSIR: \textbf{T}owards \textbf{A}na\textbf{PH}oric Ambiguity Detection and Re\textbf{S}olution \textbf{I}n \textbf{R}equirements}

\author{Saad Ezzini}
\affiliation{%
  \institution{University of Luxembourg}
    \country{Luxembourg}
}
\email{saad.ezzini@uni.lu}

\author{Sallam Abualhaija}
\affiliation{%
  \institution{University of Luxembourg}
  \country{Luxembourg}
  }
  \email{sallam.abualhaija@uni.lu}

\author{Chetan Arora}
\affiliation{%
  \institution{Deakin University}
  \country{Australia}}
\email{chetan.arora@deakin.edu.au}

\author{Mehrdad Sabetzadeh}
\affiliation{%
  \institution{University of Ottawa}
  \country{Canada}
}\email{m.sabetzadeh@uottawa.ca}

\renewcommand{\shortauthors}{Ezzini, et al.}


\begin{abstract}

We introduce TAPHSIR --  a tool for anaphoric ambiguity detection and anaphora resolution in requirements. TAPHSIR facilities reviewing the use of pronouns in a requirements specification and revising those pronouns that can lead to misunderstandings during the development process. To this end, TAPHSIR detects the requirements which have potential anaphoric ambiguity and further attempts interpreting anaphora occurrences automatically.
TAPHSIR employs a hybrid solution composed of an ambiguity detection solution based on machine learning and an anaphora resolution solution based on a variant of the BERT language model. Given a requirements specification, TAPHSIR decides for each pronoun occurrence in the specification whether the pronoun is ambiguous or unambiguous, and further provides an automatic interpretation for the pronoun. 
The output generated by TAPHSIR can be easily reviewed and validated by  requirements engineers. 
TAPHSIR is publicly available on Zenodo~\cite{Taphsir:22}.

\end{abstract}

\begin{CCSXML}
<ccs2012>
   <concept>
       <concept_id>10011007.10011074.10011075.10011076</concept_id>
       <concept_desc>Software and its engineering~Requirements analysis</concept_desc>
       <concept_significance>500</concept_significance>
       </concept>
   <concept>
       <concept_id>10010147.10010257</concept_id>
       <concept_desc>Computing methodologies~Machine learning</concept_desc>
       <concept_significance>500</concept_significance>
       </concept>
   <concept>
       <concept_id>10010147.10010178.10010179</concept_id>
       <concept_desc>Computing methodologies~Natural language processing</concept_desc>
       <concept_significance>500</concept_significance>
       </concept>
 </ccs2012>
\end{CCSXML}

\ccsdesc[500]{Software and its engineering~Requirements analysis}
\ccsdesc[500]{Computing methodologies~Machine learning}
\ccsdesc[500]{Computing methodologies~Natural language processing}
\keywords{Requirements Engineering, Natural-language Requirements, Ambiguity, Natural Language Processing, Machine Learning, BERT}

\maketitle

\section{Introduction}\label{sec:introduction}
The overall success of a project depends to a large extent on the quality of requirements~\cite{Pohl:11,Ferrari:19,Ezzini:21}. In particular, ensuring the precision and consistency of requirements is paramount for avoiding major development risks such as time and budget overruns, failure to meet customers' needs, and systems that are not trustworthy.  The requirements quality challenge is exacerbated by the fact that requirements are often written in natural language~\cite{Pohl:10}. Although natural language facilitates communication among different stakeholders, textual requirements are highly prone to ambiguity. 
At an early stage of software development, requirements engineers spend considerable time and effort  inspecting requirements specifications (RSs) to identify various quality issues such as incompleteness, inconsistency and ambiguity. 
Doing such inspections entirely manually is not only time-consuming but can also be error-prone, since engineers may overlook \textit{unacknowledged ambiguity}. Ambiguity is unacknowledged when different individuals have diverging interpretations for the same requirement, and yet, each individual is confident about their own interpretation. In such cases, the requirement from the perspective of each individual is regarded as unambiguous and thus not flagged for further discussion. 
Compared to acknowledged ambiguity that is often discussed and resolved during inspection sessions, unacknowledged ambiguity may propagate to later stages of development and lead to serious problems due to unconscious misunderstandings.

In this paper, we propose the tool \emph{TAPHSIR}, standing for \textbf{T}owards \textbf{A}na\textbf{ph}oric Ambiguity Detection and Re\textbf{s}olution \textbf{i}n \textbf{R}equirements. In Arabic, TAPHSIR means ``interpretation''. TAPHSIR focuses on pronominal anaphoric ambiguity, an ambiguity type that has been explored only to a limited extent in  requirements engineering (RE)~\cite{Yang:11, Ferrari:18}. There are no existing tools in RE to handle anaphoric ambiguity, although this type of ambiguity is prevalent in NL requirements: it is estimated that up to 20\% of industrial requirements may suffer from anaphoric ambiguity~\cite{Rosadini:17, Ferrari:18}. TAPHSIR implements the best solution emerging from our multi-solution study of anaphoric ambiguity in natural-language requirements, published in a technical paper~\cite{Ezzini:22} at the 44th International Conference on Software Engineering (ICSE 2022). This best solution is a hybrid one, where feature-based machine learning (ML) is used for detecting anaphoric ambiguity and a large-scale language model (LM) from the BERT family is used for anaphora resolution.  

Compared to our earlier technical paper~\cite{Ezzini:22}, we present in this current paper an in-depth, practitioner-oriented description of TAPHSIR, elaborating the tool's architecture and its engineering as well as how end-users can use the tool. We further discuss the accuracy of  TAPHSIR in detecting anaphoric ambiguity (including unacknowledged cases) and resolving anaphora.

\begin{figure}
\begin{center}
\includegraphics[width=0.48\textwidth]{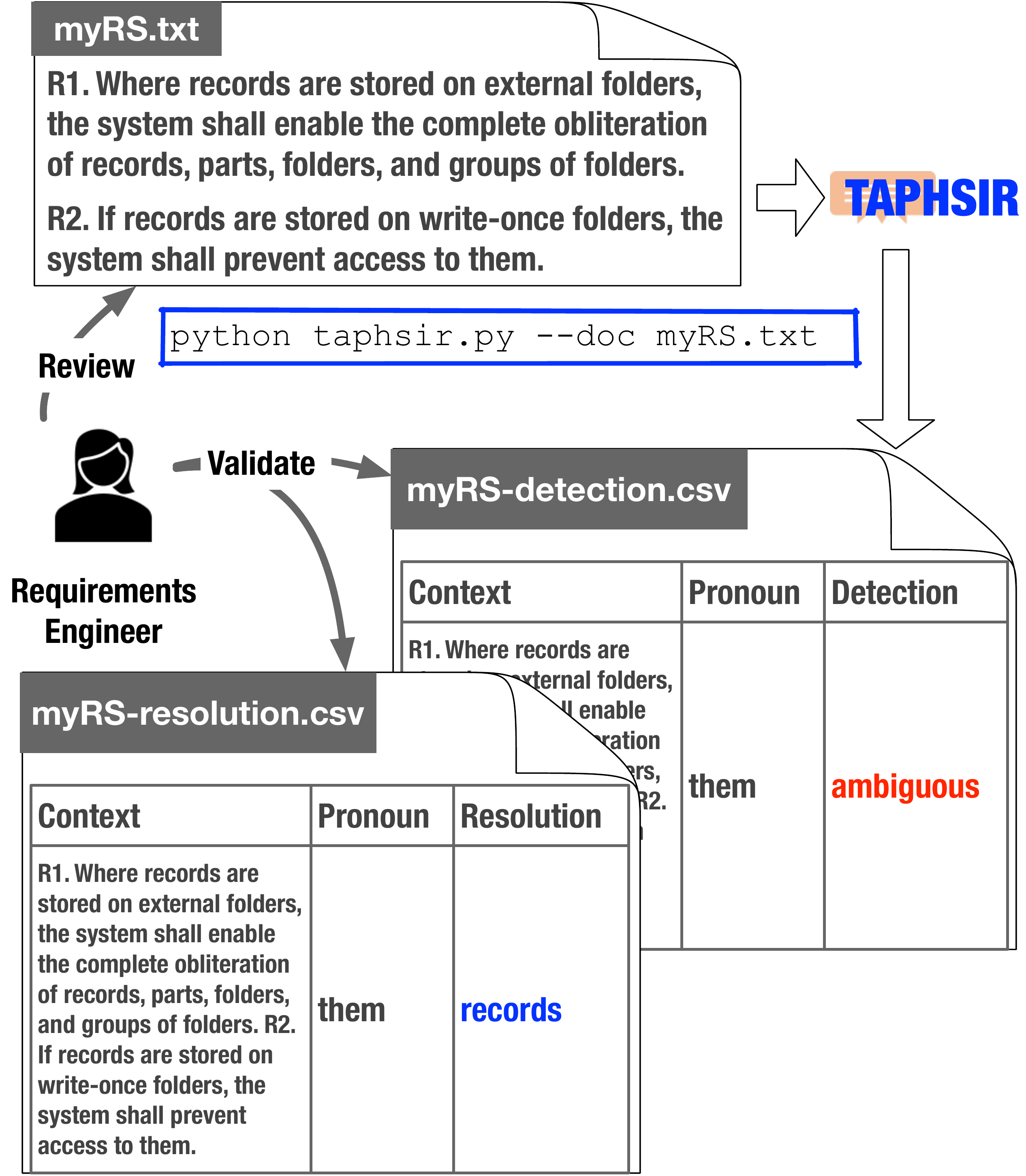} 
\end{center}
\caption{Application Example of TAPHSIR.} 
\label{fig:example} 
\vspace{-2em}
\end{figure}

TAPHSIR aims to reduce the time and effort that requirements engineers spend inspecting the use of pronouns in an RS. To illustrate, consider the  example  in Figure~\ref{fig:example}. The figure shows a requirements engineer reviewing the requirements in the file ``myRS.txt'' and using TAPHSIR for automated analysis of pronominal anaphora in that RS. The pronoun ``them'' in R2 contains anaphoric ambiguity since it is not clear whether the pronoun refers to the \textit{write-once folders} (in R2), \textit{records} only (in R1), or \textit{records, parts, folders and groups of folders} altogether (in R1). Deciding about the exact interpretation has an impact on properly implementing the requirement. TAPHSIR defines a \emph{context} for each pronoun occurrence. This context is composed of the requirement in which the pronoun occurs and the preceding requirement. Within this context, the tool identifies all noun phrases (NPs) preceding the pronoun as candidate antecedents ~\cite{Mitkov:14}. In our example, TAPHSIR will consider, in addition to those mentioned above, the following candidate antecedents: \textit{access}, \textit{obliteration}, \textit{system}. 
TAPHSIR then goes through different steps as we explain in the next section, and produces an output file (``myRS.csv'' in Figure~\ref{fig:example}). This output lists all pronoun occurrences in the input RS, and provides both the detection decision as well as the resolution result. We note that TAPHSIR can recommend a resolution also for those pronouns that are marked as ambiguous, since it applies two separate solutions for detection and resolution. Running TAPHSIR in this example requires $\approx$22.5 seconds to produce the results~\cite{Ezzini:22}.  

In the remainder of this tool demonstration paper, we outline TAPHSIR's main components. We further discuss through the lens of unacknowledged ambiguity the evaluation of TAPHSIR on a manually curated dataset (\textit{DAMIR}~\cite{Ezzini:22}).

\section{Tool Architecture}\label{sec:architecture}

\begin{figure*}
\begin{center}
\includegraphics[width=0.88\textwidth]{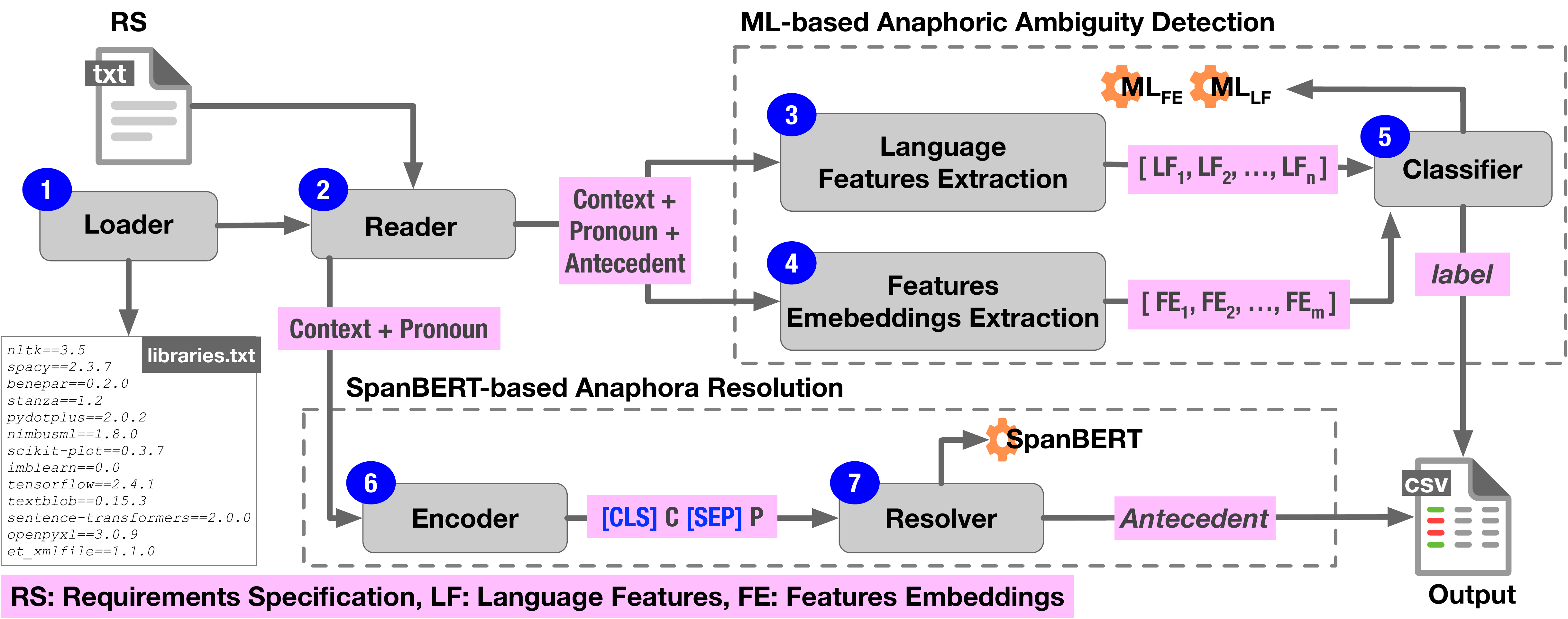} 
\end{center}
\caption{Overview of TAPHSIR Architecture.} 
\label{fig:taphsir} 
\end{figure*}

TAPHSIR is a usable prototype tool for anaphoric ambiguity handling. The tool realizes a technical solution that resulted from an empirical examination of several alternative solutions~\cite{Ezzini:22}.  
Figure~\ref{fig:taphsir} shows an overview of TAPHSIR architecture. 
The tool is implemented in Python 3.8~\cite{Rossum:09}.
Below, we discuss an end-to-end application of the tool going through steps 1 -- 7 of Figure~\ref{fig:taphsir}. 

\subsection{Preparation}
Prior to using the tool, the user needs to perform some preliminary setup. To do so, one can type in the following on the command line: 
\begin{verbatim}
python pip install -r libraries.txt 
python -m spacy download en_core_web_sm 
\end{verbatim}
The first command installs the required libraries, and the second one downloads \textit{en\_core\_web\_sm} which is needed for operationalizing the natural language processing pipeline in SpaCy. To be able to apply the tool, the user further needs to ensure that the input file is in the right format. TAPHSIR expects as input a text file (with the extension \textit{*.txt}) containing a set of requirements (or sentences). 

\subsection{Reader} 

This step parses the text of the input requirements specification, preprocesses it using an NLP pipeline, and identifies the requirements that should be subject to anaphoric ambiguity analysis. 
The NLP pipeline consists of the following seven modules illustrated in Figure~\ref{fig:taphsir}: (i) tokenizer splits the input text into tokens, (ii) sentence splitter demarcates the sentences in the text, (iii) part-of-speech tagger (POS) assigns a POS tag for each token,  (iv) lemmatizer identifies the canonical form of a token, (v) constituency parser identifies the structural units of sentences, (vi) dependency parser defines the grammatical
dependencies between the tokens in sentences, and (vii) semantic parser extracts information about words' meanings.

The output of this step is a set of \textit{triples}, each of which includes a (i) a \textit{pronoun} occurrence, (ii) \textit{context} defined as the requirement in which the pronoun occurs and a preceding requirement (recall from Section~\ref{sec:introduction}, and (iii) a likely \textit{antecedent} to that pronoun occurrence. The number of triples depends on the number of likely antecedents. In Figure~\ref{fig:example}, there are three possible antecedents as discussed in Section~\ref{sec:introduction}, namely ``records, parts, folders and groups of folders'', ``records'', and ``write-once folders''. Following this, this steps generates three triples associated with the pronoun occurrence ``them'', where each triple includes one possible antecedent. The triples will further have the same context, which combines R1 and R2.  

\subsection*{ML-based Anaphoric Ambiguity Detection} 
Our earlier work~\cite{Ezzini:22} indicates that, for the task of anaphoric ambiguity detection, (feature-based) ML leads to better accuracy than language modeling and off-the-shelf NLP methods. 
For anaphoric ambiguity detection, we employ an ensemble ML classifier that combines the results of a classifier trained over language features ($ML_{LF}$) and another trained over feature embeddings ($ML_{FE}$). 
For training and applying ML classifiers, we use Scikit-learn 0.24.1~\cite{scikit-learn}.
This component takes as input a set of triples associated with one the pronoun occurrence from the previous step, and derives as output a final label for that pronoun (ambiguous or unambiguous).

\vspace*{-.3em}
\subsection{Language Features Extraction}
This step extracts the different sets of learning features. In our work, we collected a set of 45 language features (LFs) from the NLP and RE literature. 
These features capture the characteristics of the relationship between the pronoun and its likely antecedent, %$P$ and $a_i$, 
e.g., both agree in gender or number. For extracting LFs, we use SpaCy 3.0.5~\cite{spacy}, NLTK 3.5~\cite{NLTK}, Stanza 1.2~\cite{Stanza}, and CoreNLP 4.2.2~\cite{Manning:14}. 
The result of this step is a vector representing each input triple, where each entry in this vector is the result of computing an LF. For the example in Figure~\ref{fig:example}, we will generate three vectors representing the LFs of the pronoun ``them'' and each of its likely antecedents.

\vspace*{-.3em}
\subsection{Extraction of Features Embeddings}
This step extracts the feature embeddings (FEs) for each input triple. 
FEs are mathematical vectors that encapsulate the semantic and syntactic regularities of the sentence~\cite{Jurafsky:20}. In our work, we extract 768 dimensional FEs from the BERT language model~\cite{Devlin:18}. For that, we use the Transformers library, particularly the \textit{bert-base-cased} model. Similar to the previous step, the output of this step is a vector representing each input triple. In a similar manner, this step results in three vectors for the example in Figure~\ref{fig:example}.

\vspace*{-.3em}
\subsection{Classification}
In this step, we pass the vector representation of each input triple to two pre-trained classifiers, namely $ML_{LF}$ that is trained over LFs, and $ML_{FE}$ trained over FEs. For each triple, the two classifiers independently predict a label as follows: \textit{correct} (conversely, \textit{incorrect}) indicating that the antecedent refers (conversely, does not refer) to the pronoun, or \textit{inconclusive} when the anaphoric relation cannot be inferred. 
We then apply a set of rules on the predicted labels for the triples associated with one pronoun occurrence to conclude whether the pronoun is deemed ambiguous or unambiguous by each of the two classifiers. The rules, presented in the RE literature~\cite{Yang:11}, consider the prediction probabilities produced for each possible antecedent. 

Finally, we combine in an ensemble manner the results of the two classifiers $ML_{LF}$ and $ML_{FE}$ to derive the final label for the pronoun (i.e., ambiguous or unambiguous). %decision. 
If the two classifiers agree on the label (e.g., both conclude that the pronoun is ambiguous), then this label will be the final one for that pronoun. Otherwise, the label with the highest prediction probability will be selected. This ensemble learning method yields a more accurate prediction. 

\subsection*{SpanBERT-based Anaphora Resolution}
Based on the empirical findings in our earlier work~\cite{Ezzini:22}, we know that for the task of anaphora resolution, the SpanBERT language model~\cite{Joshi:20} outperforms alternatives. Consequently, the resolution component in TAPHSIR uses a  SpanBERT model that is fine-tuned on a curated dataset from requirements. The dataset will be discussed in the next section. 
We implement SpanBERT using the Transformers 4.6.1 library~\cite{transformers} provided by Hugging Face (\url{https://huggingface.co/}) and operated in PyTorch~\cite{pytorch}. This model takes as input, from the triples generated in the first step, only the pronoun and the context in which it occurs (i.e., disregards the likely antecedents). 
As SpanBERT is originally trained to extract text spans, SpanBERT in our work predicts as output the likely antecedent for the pronoun from its context. 

\subsection{Encoder }
To be able to use SpanBERT model, the input pair of context and pronoun has to be encoded into the same format as the training data that BERT has been trained on. 
To do so, the input tuple is passed on to BERT's tokenizer which adds two special tokens: \textit{[CLS]} to represent the classification output and \textit{[SEP]} to separate the context from the pronoun occurrence. The token \textit{[SEP]} informs BERT about which pronoun occurrence to analyze in the given context.  

\subsection{Resolver}
In this step, we pass on the encoded input to the fine-tuned SpanBERT 
model and have the model predict the text span which likely represents the antecedent of the pronoun. 
SpanBERT can predict multiple such text spans with different probabilities indicating the likelihood of being the right antecedent. 
If an antecedent is predicted with a high probability (greater than 0.9), then we consider this as the resolution result for the pronoun.

\subsection*{Output} 
Given an input RS, the output of our tool is a csv file listing all pronoun occurrences in the input, and for each occurrence, providing the predicted label (ambiguous or unambiguous) and the most probable antecedent. 

\section{Evaluation} \label{sec:evaluation}

In this section, we evaluate how accurately  TAPHSIR can detect unacknowledged cases of anaphoric ambiguity and bring them to the attention of the requirements engineer.

\subsection{Dataset Description}

In this section, we use the curated dataset \textit{DAMIR} (standing for Dataset for Anaphoric Ambiguity In Requirements)~\cite{Ezzini:22}. We curated this dataset with the help of two third-party annotators who underwent half-day training on ambiguity in requirements. %To curate DAMIR, 
We collected 22 industrial requirements specifications covering eight domains including satellite communications, medicine, aerospace, security, digitization, automotive, railway, and defence.

We preprocessed this collection and prepared the list of triples (a context, a pronoun occurrence and a possible antecedent) as explained in Section~\ref{sec:architecture}. The possible antecedents for a pronoun include all of the noun phrases preceding that pronoun~\cite{Mikolov:13}. The annotators  then examined each pronoun occurrence and its possible antecedent considering the context in which they occur, and assigned a label \textit{correct}, \textit{incorrect}, or \textit{inconclusive} with the same indications as explained in Section~\ref{sec:architecture}. We then post-processed the annotations and grouped them per pronoun occurrence as follows. We mark a pronoun as ambiguous in two cases: (i) if at least one annotator acknowledges the ambiguity of this pronoun by labeling one or more associated triples as \textit{inconclusive}; or (ii) if the same triple associated with this pronoun receives different labels from the two annotators (e.g., \textit{correct} versus \textit{incorrect}). The former case implies acknowledged ambiguity, and the latter implies unacknowledged ambiguity. 

As a result, \textit{DAMIR} dataset contains a total of 737 pronoun occurrences that are analyzed for anaphoric ambiguity. About 46\% of these pronouns (342/737) are deemed ambiguous by the annotators. Out of the ambiguous pronouns, we identified $\approx$87\% with unacknowledged ambiguity, i.e., the annotators assumed that the pronoun is unambiguous yet had two different interpretations for that pronoun. 

\subsection{Results and Analysis}

To assess how TAPHSIR performs in detecting unacknowledged ambiguity, we run TAPHSIR (depicted in Figure~\ref{fig:taphsir}) on %a subset of 
\textit{DAMIR} dataset. 
TAPHSIR applies the %This component utilizes an 
an ensemble ML classifier for detecting ambiguity and SpanBERT for resolving anaphora as discussed in Section~\ref{sec:architecture}.   
On \textit{DAMIR} dataset, TAPHSIR detects ambiguous cases with a perfect recall of 100\% with a precision of $\approx$60\%, while recommends automated resolution with an accuracy of $\approx$96\%~\cite{Ezzini:22}. The perfect recall implies that TAPHSIR detects all unacknowledged ambiguous cases that were not explicitly marked by the human annotators as ambiguous.  
The precision value indicates that the requirements engineer will invest some manual effort filtering out false positives, i.e., falsely detected ambiguous requirements. In the context of ambiguity in RE, recall is often favored over precision~\cite{Berry:17}. Achieving 100\% recall ensures that all requirements suffering from all potentially ambiguous requirements will be brought to the attention of the engineers and further discussed at an early stage. 

In a practical scenario where requirements engineers review requirements under time pressure, only the requirements that are found problematic by at least one engineer would be thoroughly discussed. The engineers might not discuss those %would naturally skip the 
requirements which they could confidently interpret unaware of having multiple inconsistent interpretations. 
In conclusion, we believe that TAPHSIR has a potential in practice since it perfectly detects also those requirements with unacknowledged ambiguity which would go otherwise unnoticed during manual inspection sessions. That said, a user study is required to assess the practical usefulness of the tool.

\section{Conclusion}

We presented TAPHSIR -- a tool for detecting anaphoric ambiguity and resolving anaphora in natural-language requirements. Our current implementation reflects our findings in a multi-solution study~\cite{Ezzini:22}. TAPHSIR combines solutions based on machine learning and language models. We further evaluated how well TAPHSIR can detect unacknowledged ambiguity cases, i.e., the situation where different individuals perceive a requirement as unambiguous but, in reality, interpret the requirement differently. Our results show that TAPHSIR detects all ambiguous requirements (i.e.,  recall = 100\%) including unacknowledged cases.

In future, we plan to do a user study to assess how useful is TAPHSIR in practice. Another topic for investigation is to use TAPHSIR as a bottom layer in a broader application in analyzing requirements, e.g., using the resolution results in an extracting domain model form a requirements specification.  

\begin{acks}
This work was funded by Luxembourg's National Research Fund (FNR) under 
the grant BRIDGES18/IS/12632261 
and NSERC of Canada under the Discovery and Discovery Accelerator programs. We are grateful to the research and development team at QRA Corp. for valuable insights and assistance.
\end{acks}

\newpage
\balance
\bibliographystyle{ACM-Reference-Format}
\bibliography{main}

\end{document}